# QUINTA: Reflexive Sensibility For Responsible AI Research and Data-Driven Processes

Alicia E. Boyd
New York University
New York, USA

## ABSTRACT

As the field of artificial intelligence (AI) and machine learning (ML) continues to prioritize fairness and the concern for historically marginalized communities, the importance of intersectionality in AI research has gained significant recognition. However, few studies provide practical guidance on how researchers can effectively incorporate intersectionality into critical praxis. In response, this paper presents a comprehensive framework grounded in critical reflexivity as intersectional praxis. Operationalizing intersectionality within the AI/DS (Artificial Intelligence/Data Science) pipeline, Quantitative Intersectional Data (QUINTA) is introduced as a methodological paradigm that challenges conventional and superficial research habits, particularly in data-centric processes, to identify and mitigate negative impacts such as the inadvertent marginalization caused by these practices. The framework centers researcher reflexivity to call attention to the AI researchers' power in creating and analyzing AI/DS artifacts through data-centric approaches. To illustrate the effectiveness of QUINTA, we provide a reflexive AI/DS researcher demonstration utilizing the #metoo movement as a case study. *Note: This paper was accepted as a poster presentation at Equity and Access in Algorithms, Mechanisms, and Optimization (EAAMO) Conference in 2023.*

**Content warning:** This paper discusses sexual assault and violence and maybe triggering to readers.

## 1 INTRODUCTION

The genesis of Quantitative Intersectional Data (QUINTA) began with the need to have an honest conversation on how we frame narratives and data-centric processes, away from the dominant defaults [16]. The dominant defaults refer mainly and primarily to white, male, able-bodied, Christian, western, cis-gendered, heterosexual people. The dismantling of such dominant defaults necessitates an identification of sites saturated with power differentials that result in injustice (*e.g.,* erasure of tradition from particular communities). In doing so, the identification determines movement towards recourse. For instance, in the #metoo movement, the dominant defaults shifted slightly, from male to female via a shift in narration alongside social media virality. However, even so, an intersectional lens is required to shift narratives of power back towards communities historically marginalized and most vulnerable to erasure. Phipps [90] argued the wounds and tears of white women are centered in social movements like #metoo, which leads to the erasure of Black[1] women.

> "While white women may be positioned as victims of violence, we are also positioned as its perpetrators in relation to people of color; and that both acts and allegations of sexual violence can be used to uphold the intersecting systems of racial capitalism, colonialism, and patriarchy." [90][p. 49]

Regardless of the fact, the phrase 'me too' that turned hashtag was the life work of Tarana Burke, a Black woman, for the prior ten years (and ongoing) in harm's way [19, 57]. This paper's purpose is not to rehash the #metoo movement's ills; rather, it sets an example at how reflexivity is absolutely necessary as a way to be mindful, thoughtful in practice, and alleviate harms towards marginalized communities.

Boyd [16] documented #metoo's online viral genesis, noting tremendous focus on establishing the legitimacy of intersectionality in the #metoo. Specifically, Boyd [16] grappled with how to locate and amplify the voices of marginalized communities who experienced sexual assault and violence using traditional AI/DS techniques and processes. The disconnect between Burke's 'me too' and viral hashtag #metoo motivated, inspired, and empowered Boyd to operate differently in AI process. Therefore Boyd [16] operationalizes intersectional data-centric research praxis within the #metoo case study by critiquing the data science process. She did this by interrogating how scholars reproduced the same dynamics of the #metoo movement in their analyses of it, dynamics where cis-gended white women overtook the hashtag at the expense of sidelining those outside the white norm. Boyd [14, 16] stressed the importance of incorporating a reflexive-intersectional approach to *empower researchers* and inquiring who was included/excluded, how algorithms embed bias, and what the researcher's relationship to the project is. Likewise, this work expands on this to center the examination of the researcher's relationship to both artifact creation from data centric processes and their corresponding analyses more broadly.

In tandem with [16], AI fairness community is determined to make sense of, prevent, and alleviate existing techno-facilitated harms onto communities. An AI researchers' role and the decisions they make in their research process is not independent from the outcomes imposed onto hisotrically marginalized communities. Several works have explored how power is concentrated by those who create and design these models, yet leave the communities they impact out of design discussions [8]. In this work, we unveil this false sense of technoneutrality. It is through this unveiling of self - via a reflexive process which investigates researcher assumptions

---

[1] In this paper, we capitalize "Black" for similar reasons as expressed by the AP style guide: https://apnews.com/article/entertainment-cultures-race-and-ethnicity-us-news-ap-top-news-7e36c00c5af0436abc09e051261fff1f

(*i.e.,* via an examination of their social context, for instance) and empowers them to be aware of methods - that enables change. However, reflexivity is not thoroughly discussed; specifically – the researchers' privilege and how they get embedded into the AI and data science pipeline [16, 87].

Answering this call for a reflexive intersectional approach in data-centric processes, this work presents QUINTA. QUINTA is a methodological framework used to interrogate quantitative data-centric processes and their underlying structural and social injustices and oppressions [14]. In this work, both propose the QUINTA framework and apply it at each point of the AI/DS pipeline to help illuminate how its three core tenets are operationalized. This paper defines the process canonically as design, collection, cleaning, data exploration, modeling, and interpretation. At each step, it is discussed how researchers can be more accountable with their biases and assumptions while navigating the data centric process. Overall, this paper draws from both Black Queer Feminist and social science literature to establish reflexivity within the QUINTA framework. The contributions are as follows:

(1) Lay out and examine the foundations and discourse of reflexivity from the social science literature, explaining its benefits and critiques as it pertains to the researcher and broader knowledge production. (Section 2)
(2) Propose the QUINTA framework as one way to operationalize intersectionality within the AI research and data science (DS) community. (Section 3)
(3) Present an implementation of QUINTA via grounding the framework in a case study of the #metoo movement. (Section 4)

In this paper, there are three (3) highlighted core tenets to the QUINTA framework: the grounding of these tenets in the AI pipeline: task design, data collection, data cleaning, data exploration, modeling, and interpretation to explicitly demonstrate the exercising of reflexivity on behalf of the AI researcher. With this, reflexivity then identifies gaps and creates space for conversation on how to address and contest them.

*1.0.1 Positionality Statement.* The author is a person of color who centers an intersectional perspective in her social and professional locations and is formally trained primarily as a computer scientist. The author has additional training in Black feminist theory, gender studies, critical social theories, and queer studies, gained through activism and advocacy, which has influenced the work's posture. Further, the author is located in the United States (U.S.) but has diasporic links to other social contexts; she does her best to position this work in a global context, though to maintain scope, this work is primarily centered around the U.S. #metoo movement. The purpose of this writing is to empower individuals across the tech industry, both existing and emerging technopolicy experts, to critically flex their ML awareness through reflexive research praxis.

## 2 BACKGROUND AND RELATED WORK

In the computer science literature, the term reflexivity has been used in various areas, for instance, computer vision [79], cooperative work [23, 88], data studies [20, 34], design [42, 60, 103], human-computer interaction [43, 75, 93, 96], participatory design [42, 91], and software engineering [83]. However, the concept has been used in mixed ways and disconnected from social science literature. In this following section, we explore definitions of reflexivity and motivations for grounding oneself in a reflexive posture to more deeply center minoritized voices within the data science process.

### 2.1 Intersectionality's Call for Reflexivity for Change

Exercising intersectionality as critical praxis involves reflecting on how researchers have the power to impart their own bias, values, and social locations impact their research processes but also change their operations in order to mitigate harmful impacts. Cole [24] urges researchers to utilize intersectionality "with a new lens" of operation that goes beyond simply naming/illuminating problems. Rather, Cole pushes for the researcher to name their own biases by conceptualizing how their social position influences how they interpret their own research. As a result, engaging in a reflexive way that centers an examination of the researcher themselves is critical for the research process. Collins [27, 28] similarly calls for a critical reflection while navigating the research process in order to illuminate possible gaps and contentions in understanding within themselves. To critically reflect on one's own research practice, this includes investigating all parts of the research process such as design, data collection, methodology, metrics, and modeling. Indeed, in operationalizing intersectional research, it is critical to shift research processes away from agnostic positivist grounding and more towards socially, historically, situationally-informed praxis.

Researchers have discussed the intersections of intersectionality, artificial intelligence, and machine learning workflows (or data-centric processes), particularly critiquing how to be "fair" without causing harm to marginalized communities throughout various pipeline steps. For example, Ovalle et al. [87] challenged and explained how researchers and scientists need to operate more critically about how we do machine learning work, especially adopting an intersectionality analytical framework. These identified gaps reaffirm the need to conceptualize what operationalizing intersectionality thoroughly and holistically could look like.

### 2.2 Reflexivity

*2.2.1 Definition.* Reflexivity is an iterative, back-and-forth process that signals people to remain engaged, dissuading passivity [44, 63, 77, 107]. Reflexivity has been used throughout the humanities and qualitative disciplines, including but not limited to sociology [3, 28, 29, 38], psychology [33, 47, 51, 53, 61], nursing [32, 95, 112], ethnography [44, 64, 71], and anthropology [101]. The literature has various definitions for reflexivity [32, 36]. Webster [114] has contended that reflexivity has been poorly defined because of its complexity and fluidity of usage [68]. Nevertheless, there are many useful descriptions of reflexivity, such as "a ceaseless process" [63], "bending back" [64], and a "painful process" [38]. Barrett and colleagues [3] describes it as a "continual process of engaging with and articulating the place of the researcher and the context of the research. It also involves challenging and articulating social and cultural influences and dynamics that affect that context." [3][p. 9] Jones [63] captures its internal tensions as applied to research projects, stating: "It implicates you...is uncomfortable...[and] has got to hurt." [63][p. 124]

While formal definitions vary across scholarship, the consensus is that reflexivity's utility is to perform better research [101] and increase accountability [21, 63]. Using reflexivity in research involves interpretation and reflection [112] while being honest and open [98]. The researcher must perform a self-examination to understand themselves and their research [41], in a "dynamic that influences knowledge production." [37][p. 96] When using reflexivity, there is an integration between the researcher's personal interpretation and the creation of empirical data [2, 37]. Hence, Hand [55] urges researchers to use reflexivity at every stage of the research process, where they can examine and explicitly explain their decisions.

There is a close association between reflexivity and reflection, yet the two concepts are quite different [3, 41, 63, 113, 114]. Reflexivity is more complex than reflection due to the level of deep engagement with one's self [41]. Jones [63] compares the two concepts, arguing reflection is momentary and visceral, whereas reflexivity cuts deeper to the bone. Reflexivity involves an iterative engagement of the researcher with the research process [63, 113]. Barrett and colleagues [3] thoughtfully summarizes reflexivity and reflection: that even though people use both terms synonymously, the difference is a level of engagement and accountability that a researcher must take into account with their privilege.

Using Cole [24], she created questions or guidelines to help psychologists to think about what is missing when going through the research process. We adapt her strategy for the data science process.

*2.2.2 Benefits.* The contributions of reflexivity aid in research's "authenticity by mindfully presenting the messiness of its relational complexities." [61][p. 196] Reflexivity contains a duality which can be used as a guide for navigating the research process and challenges the assumptions and biases of the researcher [28, 32].

A major benefit of reflexivity thus surrounds how researchers hold themselves accountable by questioning and thus illuminating their biases within themselves. As Barrett and colleagues [3] argue, reflexivity is essential for the researcher given that their "own position might not always be clear to us and because we are sometimes unaware of our own prejudices and relationship with our cultural contexts and settings." [3][p. 5] Researchers are impacted and possess their own lived experiences (*i.e.,* cultures, privileges, social locations) and feelings, and their impacts on the research process cannot be ignored [64]. Incorporating reflexivity throughout the research process, as Mason [74] describes, allows the researcher to constantly be mindful of their actions and roles, and provides an additional layer of scrutiny for their research. Furthermore, Mauthner and Doucet [77] suggested there will be an increase in confidence in research when "more researchers can be self-conscious about, and articulate, their role in the research process and products." [77][p. 424]

Another benefit is disrupting habitual [69], repetitive, and performative actions [89]. Reflexive researchers are more attuned to themselves and to critiquing their techniques and decisions. MacBeth [69] describes how reflexivity begins to dismantle the power, position and privilege in the research and the researcher. This dismantling illuminates power dynamics so that researchers can see who is centered and de-centered in their work [56, 89]. Henwood [56] contends that reflexivity illuminates where marginalized communities are located and portrayed in the data - or not. Further, reflexivity provides accountability by revealing power and knowledge production, which are interwoven dynamics. The researcher and researched have an unequal relationship where the researcher has all the power. Therefore, reflexivity allows for the dismantling of this power dynamic.

Finally, when researchers participate in conversations with privilege, it allows them to contend with power implications. Chapman [21] borrows from Foucault describing how power is linked between individuals' impacts on society. "Power relations permeate our lives, and simultaneously power passes through individuals to renegotiate or perpetuate social structures." [21][p. 725] Therefore, this establishes the need for a reflexive posture on how we perpetuate biases in data analysis and the tools we utilize to perform it.

Chapman [21] further points out, "often people from dominant groups occupy these institutionally sanctioned positions of dominance, to a degree that is statistically not representative of a population as a whole." [21][p. 725] Although Chapman comes from a sociological lens, their points still apply to other domains and disciplines, including computer science, where reflexivity allows us to understand how power flows through the tools and technology we use to analyze data, and to re-conceptualize "how to resist or navigate out our own involvement in the systemic oppression." [21][p. 725]

As it pertains to AI/DS processes, as ubiquitous adoption of data-driven technologies continue, a growing call for ways to grapple with critical inquiry and critical praxis that attends to social justice and reducing algorithmic harms is ever present. As such, reflexivity finds itself well situated as a form of expanding algorithmic fairness towards more holistic critical praxis.

*2.2.3 Critiques of Reflexivity.* While reflexivity offers these benefits, it comes with several criticisms. One disadvantage is that it is not a straightforward process [36]; depending on the constraints of a project, it might be too time consuming to faithfully implement. Reflexivity may reveal various nuances and roadblocks that were not originally accounted for, such that researchers might not have the flexibility to complete the reflexive process. Researchers have proposed guides and documentation to alleviate implementation burdens, and recommended transparency about any obstructions to reflexive engagement, if it cannot be completely implemented.

Another critique when using reflexivity are the dangers of narcissism, self-righteousness, and nihilism [113]. Reflexive researchers can develop savior complexes, and the process can become a vehicle for excessive confessionalism [41], which can deviate from a project's core goals. Webster [114] argues researchers should be explicit why they are using reflexivity, and not use it to center themselves. Being explicit allows researchers to pre-set boundaries, so they will not become blurred later on [107].

When scholars have employed reflexivity, they expressed the challenges of engaging with the concept, mainly calling attention to their power and privilege. For example, Jones [63] discussed his rumination for putting reflexivity into practice while embattled by white privilege even though being a member of the queer community.

> "Reflexivity has got to hurt. Reflexivity is laborious. But while it may be laborious for me to go out of my way to intervene and how I perform privilege, I must also recognize that it is a privilege to not have my performance always already marked as marginal." [63][p. 124]

Jones's description of engaging with reflexivity in his work draws attention to three things: pain, difficulty, and the constant reminder of his privilege while doing so. When reflexivity is implemented, it will not feel good. Jones [63] describes reflexivity cutting to the bone, where there is a consequence of implicating the researcher's involvement in power and privilege [63]. It is difficult to not give up from the pain and fall into the habitual habits because it would be easier. Although exhausting, this work is necessary to be conscientious of impacts on vulnerable populations.

Finally, a criticism of reflexive research stems on whether it can be applied in the quantitative domain. Reliability, rigor, and validity are concerns when researchers use reflexivity [105], though scholars still encourage its use [52]. Prior to 2016, reflexivity was less common in quantitative research [37]. Some were hesitant due to concerns about the validity of control measures [92, 113]. Others [32, 53] have since pushed back on this notion, arguing that quantitative research is not devoid of personal values and biases, which have implications on data collection, measures, techniques, and interpretation [112]. Furthermore, Darawsheh [32] argues that reflexivity should absolutely be used, especially interrogating measures used to influence how we think about people and things. Scholars have criticized quantitative research for minimal acknowledgment of the researchers' positions and hidden assumptions [78, 99]. Ryan and Golden [99] argued that a "reflexive approach would not undermine the value of the research study but would add a depth of understanding about how, where, when and by whom data were collected." [99][p. 1198] Reflexivity brings transparency and provides "information about the positionality and personal values of the research" [112][p. 38], facts that have real implications for the quality of quantitative research. Guillemin [53] contends "The goal of being reflexive in this sense has to do with improving the quality and validity of the research and recognizing the limitations of the knowledge that is produced, thus leading to more rigorous research." [53][p. 275]

This work attends to these critiques by creating a reflexive framework that abides by a quantitative space and allows for researchers to ask questions throughout their quantitative methodologies. They are able to explore new forms of transparency and positionality so that the quality of work, especially as it pertains to data-centric processes, is both improved and re-centered towards marginalized voices.

### 2.3 Existing AI Fairness Reflexivity Frameworks

A few scholars have encouraged the use of reflexivity for researchers to attend to power and accountability through specific steps of the data science process, such as documentation [79] and data creation [34]. Within the AI fairness community, reflexivity has been discussed in documenting computer vision datasets [79] and assessing the gaps and contributions in the FAccT community [66]. However, in this paper the focus is on providing a framework (1) grounded in reflexivity through intersectionality and (2) ground in a case study with respect to the metoo movement.

Importantly, transparency is key to engaging reflexively but it does not entail it. Scholarly works pertaining to transparency are necessary to operationalize just AI at various parts of the AI pipeline. However, copious literature only seems to focus on transparency and scrutiny on the particular outcomes themselves, rather than a procedural reflexive analysis throughout. Importantly, [46] and [80] center the examination of ML artifacts. However this work differs in that this framework directly shifts the research gaze towards that of marginalized identities through the examination of the researcher their-self.

With respect to more socio-centric analyses, Birhane et al. [9] critiques existing participatory approaches in ML processes, which is absolutely needed as technologies seek to understand their needs more. Suresh et al. [108] also explores centering communities and what the researcher to stay mindful of throughout ML design. However, both do not does not go into depth about reflexivity. For example, Birhane and Guest [8] centers participatory research, rather than the researcher their-self. In this work, QUINTA proceeds with centering an examination of the normative reasoning behind research approaches in AI/DS pipelines. QUINTA empowers the researcher to be be reflexive. A research team may not have the necessary resources to include and incorporate participants in the AI/DS lifecycle. QUINTA contributes to this space by providing a reflexive guide for researchers (or a team of researchers) to navigate the AI/DS processes from beginning to end. In this work, reflexivity is being used similarly in that both of the questions are meant to serve as a guide, not an exhaustive checklist, rather as a tool to help guide them through a reflexive exercise. QUINTA focuses on the researcher, not the participation of others (*i.e.,* the community); the situatedness of the researcher is different. QUINTA is a solo or group effort not involving the community. If it does involve the community, then refer to this paper. In the end, we all need all the guides to help navigate ML processes.

## 3 REFLEXIVITY FOR SOCIAL JUSTICE

The inspiration to use reflexivity in QUINTA stems from two places: (1) the benefits discussed earlier, and (2) how scholars have used this concept jointly with intersectionality. When reflexivity and intersectionality are used intentionally, there are three themes that arise. One, there are conversations surrounding themes of power. Two, the researcher and research process are repositioned towards the margins. Three, social justice[2] becomes a part of the conversation.

---
[2]Scholars argue that justice is an essential attribute of intersectionality which should not be omitted from the discussion [12, 28, 94]. Collins [28] calls out, today's intersectional projects frequently "do not deal with social justice in a substantive fashion, yet the arguments that each discourse makes and the praxis that it pursues have important ethical implications for equity and fairness" [28][p. 47]. Historically, justice was such an outwardly centralized part of intersectionality that there was little need to "examine it or invoke it" [28][p. 47]. This illuminates a disconnect between intersectionality and how the theory is used in varying venues. By contrast, other scholars [1, 63] have made connections between intersectionality and justice to challenge their own privileged positions of power in their work, calling for justice as an explicit component for doing so. In particular, Adams [1] argues for justice when using intersectionality, on the

Collins [28] contends that reflexivity is already integral to intersectionality: "Intersectionality already has within its theory and practice an expansive set of ideas and practices that enable it to be self-reflexive and accountable." [28][p. 64] Yet, Collins goes on to express concern: "How self-reflexive are intersectionality's practitioners about their own inquiry and praxis?" [28][p. 64] The accountability gap is made explicit in the QUINTA framework in order to reckon with researchers not intentionally critiquing their power and its implications on praxis. Moreover, intersectionality is not an "a la carte" paradigm; it is weakened and destabilized when not used in its totality. Therefore, with the QUINTA paradigm reflexivity and intersectionality are used together, which further, explicitly calls out reflexivity to promote the critical assessment of the researcher's privileged position in the research process, as its generator or creator. QUINTA is intended to implement Bowleg [10, 12] and Salem [100]'s precautions for being mindful of how intersectionality is used, and for avoiding corrupting the methodology from its roots in Black Queer Feminism.

## 4 WHAT IS QUINTA?

QUINTA is a methodological framework that emphasizes intersectionality and reflexivity as foundational concepts for supporting research scientists' ability to evaluate their data and techniques for representation and potential bias while engaging in critical reflection on their practices. Figure 1 illustrates how reflexivity, intersectionality, and the data science process interact to form QUINTA. The illustration's spiral representation was chosen because it embodies reflexivity as an iterative and ceaseless process, looping back on itself at each step. Intersectionality, for its part, is the milieu within which reflexivity operates; like a jellyfish and water, reflexivity loses its coherent structure without intersectionality.

Reflexivity and intersectionality thus work together within the QUINTA framework, and the approach is integral to its application. Researchers interrogate each methodological choice from a reflexive-intersectional perspective at each stage of the process. QUINTA is applied by iterating a set of questions over the data science process until one arrives at a suitable methodology. The questions are formulated to guide and reveal who is being centered and who is being marginalized or erased. They provide an iterative vehicle for illuminating whether the researcher inclusively meets the goals and whether those choices or decisions are inclusive.

Through this application, QUINTA can reveal and contextualize biases that would otherwise go unremarked, or worse, be presented in a "view from nowhere" manner that mistakes a lack of obvious or intentional bias for proof against the possibility of bias. QUINTA does not itself guarantee that any result will be inherently unbiased; rather, it is the exercise of contextualizing and highlighting bias, which allows researchers to address it. As Boyd [16] outlines, this approach is incorporated as three broadly defined questions about the power relations at each step of the data science process, seen in Table 1.

---

basis of its roots therein. Moreover, scholars trace their intersectionality to its Black Feminist Queer roots when positioning it with reflexivity and justice. Per Bowleg [12], when authors cite "the trinity" (Crenshaw [30, 31], Combahee River Collective [25], Collins [26–29]), this authentically captures its origins in lived experiences.

Referring to Cole [24], her grounding was using intersectionality to develop these questions. She did not explicitly use the term "reflexivity" but what she was doing was being reflexive in this process. The first question addresses who is included and/or excluded from the data. This question can lead us to look deeper into where and how this data was collected, from which sources our data's biases might originate, and its harmful impacts of various technologies, tools, and techniques on communities and groups. These impacts can maintain and perpetuate stigmas, and reinforce and create new biases, so it's important to find them. Finally, the third question engages with the conceptualization of reflexivity. These questions will be explored through the next three subsections.

### 4.1 User Inclusion

**Who is included in the data?** At the most basic level, researchers should consider who is included in the data they collect, as this question draws attention to who has been included or excluded in the collection process. Because certain groups or communities have historically been excluded or hyper-focalized, there is a tendency to extend generalizations without considering whether they apply to other populations, nor whether those populations are even present. Therefore, it is essential to interrogate the source of the data and the particular people it focuses on.

This question also illuminates who is being overlooked and who is privileged. The need for data representation was well-illustrated by early work showing that a single-axis framework privileges some, whereas it creates erasures for other groups [24]. As Cole [24] discusses inclusion, it "transcends representation, offering the possibility to repair misconceptions engendered by the erasure of minority groups and marginalized populations." [24][p. 172] By starting our analyses on the margins, we encompass the understanding of experiences on the fringes, rather than encoding dominant, implicitly assumed perspectives into our datasets. The goal is to disrupt these assumptions by including multiple identities. Such attention is critical, because failure to include these identities reinforces biases and perpetuates stigmas. In addition, the act of inclusion itself helps scientists to ask further questions about how the data is collected, and how to maintain that inclusivity [40].

### 4.2 Methods

**What role/How does ML/AI/statistics embed and amplify inequality?** A key lesson of intersectionality is that statistical methods cannot simply produce neutral, unbiased results by quantifying and subtracting bias [11, 40]. All numerical methods inherently encode historical and continued relations of the political and material, of social inequality and stigma [104]. Understanding these shortcomings, QUINTA can critique and interrogate algorithms, tools, and techniques. ML and AI are typically represented as embodying unbiased normative ideals, but several scholars [65, 85] have argued that these tools and techniques cause harm to marginalized and vulnerable communities. Proposals for mitigating these harms have ranged from FAIR and FATE [3], to outright abolishing these tools. While FAIR and FATE [6, 45, 62, 82, 102] superficially appear progressive by enacting "safeguards" to hold algorithms accountable,

---

[3] "fairness, accountability, transparency, and ethics"

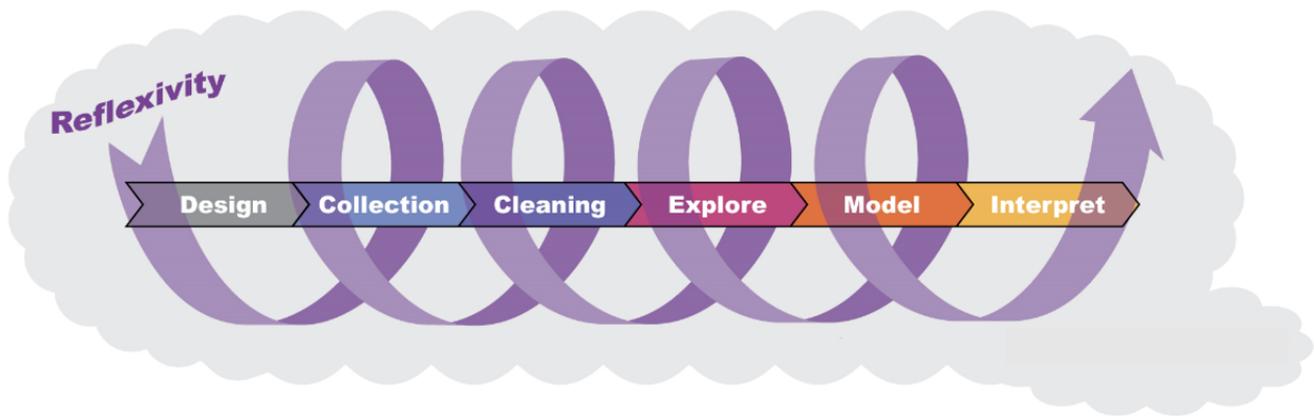

Figure 1: QUINTA Diagram

| Phase | Who is included in the data? | What role/How does ML/AI/statistics embed & amplify inequality? | What is your position in relationship to the data? |
| --- | --- | --- | --- |
| Design | Is the workflow attuned to diversity and inclusion? Did the literature review attend to the social and historical contexts of inequality of the social issue(s)? | Did the literature review attend to the social and historical contexts of inequality of the proposed techniques? | Why are you doing this research? What is the personal benefit for you? Are you causing harm and erasure? What story are you trying to tell? |
| Collection | Have you thought about neglected groups? Who are you centering and who are you marginalizing? | Do the data sources used exacerbate unequal visibility and further enhance structural inequality? | Who is included or excluded from the collection? Why and how are you silencing or amplifying them? |
| Cleaning | Are normalizing and cleaning techniques reinforcing a dominant reframing or are they promoting inclusivity? | Do the techniques used exaggerate unequal visibility and further enhance structural inequality? | Who are you silencing/amplifying by cleaning the data? Why is that happening and how can you fix it? |
| Explore | Are you developing and investigating adequate measures for using the data? | Are differences conceptualized as stemming from structural inequalities produced from the techniques? | Who are you silencing/amplifying in the models you are exploring? Why is that happening and how can you fix it? |
| Model | Does your evaluation attend to diversity and inclusion in the data and outcomes? | Are you testing for both similarities and differences in outcomes for different groups (e.g., implicit bias)? | Who are you silencing/amplifying in your model selection (e.g., representation)? Why is this the best model and if it is not how can you fix it? |
| Interpret | Are findings being interpreted to represent a universal or normative experience? | Are you considering how similarities and differences in outcomes are interpreted and how the structural inequalities are enhanced by the algorithms and statistics? | Who are you silencing/amplifying in the interpretation of your model outcomes (e.g., sensitivity to nuanced variations in the data)? |

Table 1: The overarching and phase-specific questions driving reflexivity intersectional at the foundation of QUINTA [14].

the problem is that these concepts are centered from a dominant positionality [5, 45].

Recent work around fairness, ethics, and accountability has called attention to various inequalities (*i.e.,* health care, finances, loans, *etc.*), but only does so temporarily. These ideas fall short because they are centered on and around dominant positions, not addressing the inequities of those on the margins. The real question is, how can we become intentionally inclusive and promote equity when we continue to use tools and techniques steeped in the marginalization of communities? Scholars have highlighted concrete examples of data science tools harming queer, non-able, and neuro-diverse communities [54, 65, 97]. By asking how our tools embed inequality, we can choose techniques and tools that consider multi-dimensional and overlapping social identities, and how systems and structures of oppression impact them.

### 4.3 Researcher position to data

**What is your position in relationship to the data?** The concept of reflexivity is integral to this third question. By being engaged and conscientious at each stage of the data science process, researchers can critique and question their decisions and results. This question thus spurs the iterative reflexive process, encouraging practitioners to embrace discomfort and evaluate their own position in relation to the data. This adds a critical layer of accountability to research.

## 5 CASE STUDY: #METOO

For this section, each prompt will be taken from QUINTA (see Table 1) and apply it to the work that Boyd [16] did.

After discovering Tarana Burke's work a decade ago, it became apparent that other communities needed to located. Are they on Twitter? The objective was to identify marginalized communities who have experienced sexual violence. Thus, it initiated data collection from two points of reference: Alyssa Milano, who popularized the movement, and Tarana Burke, who did not actively engage on social media due to the nature of the conversation, resulting in a lack of previous dialogue on the platform.

Gaining insights from Tarana Burke's knowledge, the intention was to assess how people discuss the #metoo movement from both ends. Sexual assault and violence is not a singular issue, but the misconception was that it was solely a concern for white individuals. However, it is not limited to white communities. The movement provides communities with power, amplifying their voices and offering hope, resistance, and a sense of existence. It aims to resist and change the prevailing narrative, striving to be seen, heard, and believed. Historically, these populations have been overlooked, silenced, and disbelieved. Tarana Burke provided them with a platform to openly discuss their experiences, aiming not to portray them as victims but as survivors. It created a safe space for open dialogue and facilitated a healing process described as "empowerment through radical healing." This perspective diverges from the dominant media narrative, shedding light on the voices that often remain unheard. Healing emerges through accountability and visibility, healing in community, and empowerment achieved through stepping outside the confines of relying on a dominant group for validation. How individuals navigate healing and hope on their own terms becomes a central question.

### 5.1 Task Design

**Is the workflow attuned to diversity and inclusion? Did the literature review attend to the social and historical contexts of inequality of the social issue(s)?** To uncover underrepresented voices, the process was initiated by collecting data on the conversations surrounding Milano and Burke in relation to the #metoo movement. Then, community detection algorithms were employed to identify emerging communities based on the hashtags used in these discussions. It was observed that the #metoo hashtag was evolving, reflecting its diverse morphological nature. In particular, we noticed that under Tarana's tweets, there was a greater focus on marginalized communities compared to Milano's tweets, which predominantly represented mainstream media and had fewer variations. Milano's tweets displayed a significant level of communal homogeneity, echoing what was portrayed in the media. Conversely, Tarana's tweets exhibited greater diversity, with a higher level of heterogeneity in the hashtags used under Burke's name. Collecting tweets using the #metoo hashtag alone would only capture the dominant discourse, which mirrored published articles and social media more broadly. However, by focusing on Tarana's tweets, conversations were accessed on the fringes, where communities existed on the periphery and engaged in more personal discussions. Therefore, specific metrics had to be devised that were relevant to Tarana's conversation, such as measuring the number of derivatives that emerged.

The research project's workflow was designed to create two networks (see Figure 2): a QUINTA network and a non-QUINTA network. The latter employed a generic approach where the lower-degree nodes were eliminated to give more attention to the main communities in the dataset. The former followed the reflexive-intersectional approach, removing the larger-degree nodes within the dataset to reveal smaller communities. The intent was to extend beyond the dominant racial and gender lens of the #metoo movement.

The literature review found ample historical and social context relating to inequalities in prior #metoo movement scholarship. Both the network analysis [106, 115] and synopsis perspectives portray the movement as unifying and unified, yet more focus on white cis-hetero affluent women narratives despite the fact that sexual assault and violence impact all social and identity positionings [15, 49, 59, 70, 76, 86, 90, 99, 109]. Furthermore, given the numerous critiques of co-option from the original 'me too' movement [4, 15, 35, 48, 50, 76, 86, 109], the design of the network analysis workflow needs to capture the hidden users in the movement.

**Did the literature review address the social and historical contexts of inequality of the proposed techniques?** To fully comprehend the inequality surrounding the #metoo movement, as well as issues of sexual assault and violence, it was crucial to consider the social and historical contexts. Each marginalized identity affected by these issues has a well-documented but often overlooked history of exclusion when compared to the dominant identity. During our examination of previous scholarly works [106, 111, 115] that employed network analysis to study the #metoo hashtags, we identified a gap in identifying intersectional communities within the #metoo discourse. Notably, Xiong et al. [115] and Suk et al. [106] argued that the data sets contained intersectional communities actively participating in global online conversations, yet these communities were disregarded or absorbed within the larger analysis. This is significant because numerous scholars have expressed concerns about how this absorption has undermined the grassroots nature of the original #metoo movement and obscured the voices of marginalized intersectional individuals. The proposed QUINTA technique aims to address these gaps and rectify these issues.

**Why are you doing this research? What is the personal benefit for you? Are you causing harm and erasure? What story are you trying to tell?** The #metoo movement is one of the most vivid and widely-recognized examples of an online social movement. A popular narrative about the history of #metoo is that the movement started on October 15, 2017, in response to a viral tweet from celebrity Alyssa Milano. However, the anti-harassment social movement #metoo actually began in 2006 when activist Tarana Burke wanted "to provide space for women of color to understand that they have a deep worthiness just because they exist" [66]. As Milano's tweet went viral, Burke's contributions were at risk of being ignored or co-opted. Sexual assault and violence do not happen only in Hollywood; they are unfortunately ubiquitous. Some

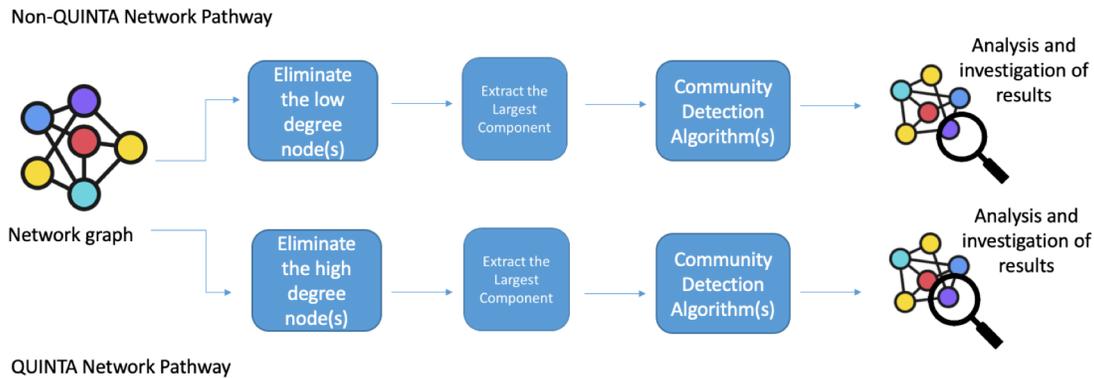

Figure 2: QUINTA Design in #MeToo Case Study

survivors[4] of sexual assault and violence possessed multiple social intersections of race, gender, orientation, able-bodiedness, class, nationality, religion, *etc.* which were also at risk of being ignored. It is their story that we are trying to tell.

The viral nature and popularity of an online movement provided researchers with a significant opportunity to study its societal and communal impacts. However, existing scholarship on #metoo largely neglected survivor communities beyond Hollywood and predominantly focused on Milano [72, 73, 84, 115, 117]. Many #metoo researchers primarily centered their analysis on the experiences of white women, mirroring the narratives prevalent in news media, which entered on the cis-gendered, affluent, able-bodied, white women in Hollywood. Consequently, their findings and conclusions failed to capture the intersectional nature of the actual #metoo movement.

Accordingly, this study's design is intended to address these harms and erasures. Other scholars [15, 22, 67, 86, 110] illuminated this disconnect and called attention to the question of who was getting centered. Boyd and McEwan [15] specifically argued that the sudden increase in the #metoo hashtag's visibility paradoxically led to the erasure of the Black female and LGBTQAI+, voices that had initially established the 'me too' movement, thus causing harm to those communities. Boyd [16] expanded this investigation into quantitative spaces, particularly data science. The underlying question was how data scientists and researchers could mitigate these biases to prevent further harm to vulnerable communities via our methodological approach, to be more consciously aware of and lessen the harm our work causes. In answering this question with QUINTA, our motivations and personal benefit are one and the same: to recenter the #metoo research on marginalized voices.

### 5.2 Data Collection

**Have you thought about neglected groups? Who are you centering and who are you marginalizing?** Previous data science scholarship researching the #metoo movement mostly used a garden hose and random sampling methods to collect data for network analysis. In addition, Trott [111] observed that intersectional narratives were absent from the first day and throughout, demonstrating how popular white feminists dominated the protest's core. Yet, Trott [111] found that Tarana Burke was located on the periphery of her network analysis. Because of all these factors, the motivation to use snowball sampling (see Figure 3) is where the researcher picked the two seed points of Burke and Milano. These two women were selected for their unique roles in the #metoo movement as, respectively, (1) the creator of the 'me too' movement and (2) the celebrity widely credited for the 'me too' phrase and hashtag going viral on Twitter. In essence, this is inherently an act of centering. The purpose of this centering is to allow the snowball sampling to reveal groups that would be neglected in the garden-hose approach.

[4]We use the term "survivor," which is what Tarana Burke [18, 19] used to empower those who have experienced sexual assault and violence.

The data collection needed to implement a sampling technique that encapsulated the communities formed around the movement's two focal women, Burke and Milano. The researcher used snowball sampling [7, 17] to collect the data from Twitter to see who is participating in this movement solely from these two women's involvement and capture the intersectional population involved in #metoo. This collection method also avoids high-engagement users who are often over-represented in garden-hose techniques, thus increasing the visibility of marginalized identities. This coincides with researchers [13, 39, 40] who emphasize the use of methods that are suited for finding "marginalized populations who are not readily accessible using more traditional sampling methods such as random sampling" [13][p. 338] and garden-hose samplings. To summarize, in addition to the underlying marginalizations and inequalities that the snowball sampling has been designed to center, the over-representation of high-engagement users is another form of structural inequality that we have taken care to address.

**Who is included or excluded from the collection? Why and how are you silencing or amplifying them?** Since collection needs to be mindfully designed to incorporate maximal heterogeneity, we proceeded with the snowball sampling technique. Under the snowball sampling scheme, there were two inclusion criteria to collect a Twitter account in a "generation," or iteration of sampling: (1) having been "@"-mentioned by a user in the previous generation, and (2) having used the hashtag #metoo (case-insensitive) in a tweet. As a result, any users who were not part of this network were excluded, which allows for the extreme amount of "noise" associated with garden-hose techniques to be filtered out, allowing for marginalized populations to be detected.

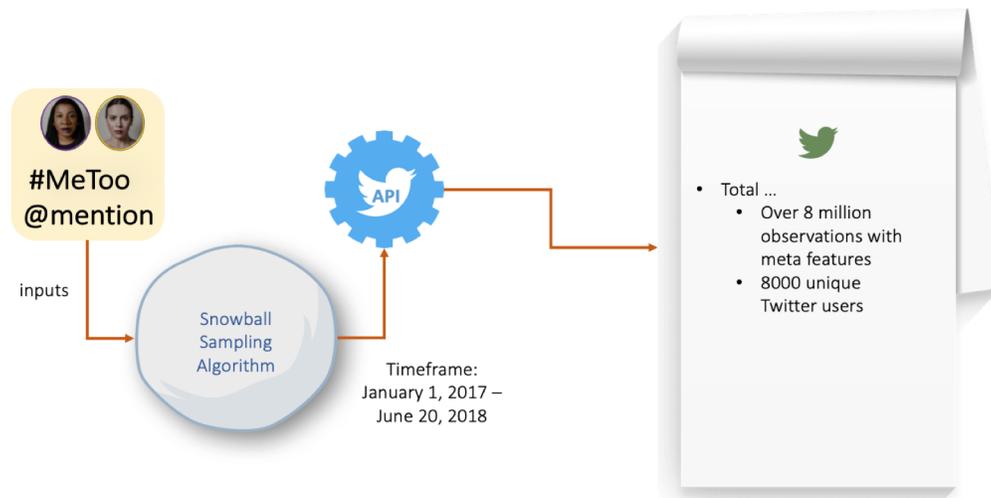

Figure 3: QUINTA Collection in #MeToo Case Study

### 5.3 Data Cleaning

**Are normalizing and cleaning techniques reinforcing a dominant re-framing or are they promoting inclusivity?** While cleaning the data, the focus scope was on English written tweets. A limitation was excluding non-English tweets due to the presence of non-ASCII characters. This limited the context to primarily focus on the #metoo movement in the United States and the broader anglophone world. This includes people who speak some English but for whom it is not their primary language. In the American context, this centers the dominant language group, which overlaps with the identities of the two seeds, but excludes speakers of non-dominant languages. Globally, English is dominant as a lingua franca, but is not the only dominant language.

**Do the techniques used exaggerate unequal visibility and further enhance structural inequality?** Focusing on English only tweets limits the conversation and the expansiveness of the #metoo hashtag. The scope is limited to how the hashtag shifted into contexts outside of the english-language domain. Therefore, there was a great attempt to use the location metadata. Unfortunately, it was unavailable, but our assumption is that the majority of non-English tweets were not from the U.S., with the possible exception of some appreciable proportion of Spanish-language tweets, given its status as the second-most popular language in the U.S. The exclusion of non-English tweets renders them invisible to this analysis. Scholars [81] documented how the #metoo hashtag was transformed differently within Japanese culture, for instance. This analysis will missed out on opportunity to examine non-english tweets. Unfortunately, it was hard to avoid the exclusion of non-English tweets given how global the viral hashtag traveled across the world.

**Who are you silencing/amplifying by cleaning the data? Why is that happening and how can you fix it?** The largest subset of non-English tweets were in East Asian languages; thus, East Asian voices were particularly impacted by silencing in this data cleaning. Likewise, the Spanish-speaking American population was also likely impacted. Due to their large proportion of the American population, this is not insignificant to the American context of the study, but it may be a mitigating factor that this population is highly bilingual in English. Other non-English-speaking groups bear mentioning as having been silenced as well.

### 5.4 Data Exploration

**Are you developing and investigating adequate measures for using the data?** At first glance, using metrics like retweets and favorites may seem like a approach to analyzing the data with respect to the objective of identifying community stories besides the dominant narrative. However, only doing this does not lend itself well to capturing heterogeneous voices that may be of smaller size. Therefore, we explored what impact two major types of hierarchical [5] clustering, divisive and agglomerative, have on the data.

For the investigation of adequate measures, community detection algorithms are used. Since community detection algorithms are often underpinned by maximizing the likelihood of node presence with respect to node degrees, we acknowledge that doing this with larger communities—when we want to examine small communities—may run the risk of overshadowing them. Therefore, we try to use algorithms that focus on detecting communities for smaller networks. For instance, the Edge-Betweenness algorithm is a divisive type of community detection algorithm that identifies communities in smaller networks [116], placing more emphasis on the importance of a node within its network. By contrast, the WalkTrap algorithm employs short "random walks" to identify communities within networks, where the walks remain confined to the same community [116]. By using two algorithms with complementary focuses, we increase our chances of detecting significant information.

---
[5]The word 'hierarchical' implies there is a rank (or order) to how objects are grouped together.

**Are differences conceptualized as stemming from structural inequalities produced from the techniques?** We centered on conceptualizing differences based on structural network qualities. For example, undetected structural inequalities may be produced during the elimination of nodes from the network graph. However, the types of hashtags we see when nodes are eliminated from each step depend on the technique used, which is why multiple techniques were used with complementary focuses. The overall effect is a structurally more equitable view of the entire data set.

**Who are you silencing/amplifying in the models you are exploring? Why is that happening and how can you fix it?** In order to identify silencing or amplification in the models, several models and measurement outcomes were tested. Creating different pathways and testing several modes of node elimination (Figure 4) illuminated who is being silenced or amplified by each technique. We hypothesized that if we remove lower degree nodes (in the Non-QUINTA Network Pathway), this will center on more popular (and thus likely dominant) communities related to the #metoo hashtag; whereas in the QUINTA pathway, when we remove the highest degree node (which is #metoo), we may see an increase in community heterogeneity which increase abilities to capture smaller and more marginalized communities. By using selective silencing and amplification within each technique, and using both techniques together, their complementarity "fixes" each other.

### 5.5 Model

**Does your evaluation attend to diversity and inclusion in the data and outcomes?** At first glance, using metrics like degrees, retweets, and likes is a good approach to analyzing the data. However, only doing this does not lend itself well to capturing the voices of those who may be smaller in size. Therefore, to address attention to diversity and inclusion in the data and outcomes, metrics were developed that capture community heterogeneity as a means to consider modeling quality. Furthermore, it was intentionally chosen to split across a QUINTA and a non-QUINTA pathway in order to contrast research outcomes when reflexivity is engaged. In doing this, we observe minor differences between the two community algorithms within each respective pathway (see Figure 5). However, there were stark differences in node removal when comparing the two pathways against each other. Depending on which pathway was taken, the non-QUINTA would not be oriented toward diversity and inclusion based on the hashtags present. In the QUINTA pathway, more hashtag derivatives were present, representing intersectional communities. In exploring these differences across the pathways with respect to how they define both outcomes and collect data, they attend to more inclusive praxis.

**Are you testing for both similarities and differences in outcomes for different groups (*e.g.,* implicit bias)?** A grounding contrast is made by highlighting similarities and differences in this task by splitting the research process into one that centers on dominant voices versus one that does not. Furthermore, it was intentionally chosen to analyze different outcomes with respect to how the removal of a dominant voice shifts conversation heterogeneity via hashtag derivative measurement. In particular, various outcomes were tested, such as the removal of nodes with lower weights and the removal of the hashtag #metoo, as it represents a consuming, monolithic conversation that overshadows others around it. In the non-QUINTA pathway, when nodes with lower weights were removed, it gave more priority to nodes that were more central to the network. As a result, lower nodes represented less frequently used hashtags. When removing the #metoo tag from the non-QUINTA pathway, it was found that even if the pathway did not have the #metoo hashtag, it overshadowed mostly everything else. Meanwhile, when the most monolithic node, the #metoo network, was removed in the QUINTA pathway, it led to more visibility for the other nodes on the network's fringes.

**Who are you silencing/amplifying in your model selection (*e.g.,* representation)? Why is this the best model and if it is not how can you address it?** Model selection influences how intersectional communities are exposed, amplified, or go otherwise unseen. Therefore, the implementation of the network analysis involved creating two strategies. The first focused on the #metoo hashtag revealing monolithic conversations involving sexual assault and violence. The second involved removing #metoo to reveal nuanced community-specific implications of sexual assault and violence. Emerging themes arose from each strategy (1) more general conversations surrounding the main hashtag and (2) removing the main hashtag exposed intersectional communities identified by hashtag derivatives. Each strategy uses the concept of silencing - removing the main hashtag, or the main hashtag silencing others - to detect different communities, but it is clear that the second strategy is superior in detecting marginalized communities.

The implementation of network analysis was driven by the ability to visualize the various hashtag communities in a network. In each pathway, community detection algorithms were used that are hierarchical in their clustering of similar objects. Furthermore, choosing *not* to follow the dominant practice of removing low-degree nodes for modeling to prevent the erasure of various community conversations. In using these algorithms, it is evident that there is a future need to investigate the mathematics of each community detection algorithm to determine if any inclusive or exclusive aspects are overlooked.

In the QUINTA pathway, we are led to travel to the margins of the data instead of focusing on the most monolithic node - that is, #metoo. Therefore, eliminating the #metoo node from the network gives more visibility to the other nodes. Removing the largest node, #metoo, reduced the node strength range from 0 to 18. Additionally, this resulted in a drastic reduction in the number of edges, from 83 to 65. The rationale behind intersectional network methodology is the ability to visualize the nodes that are not readily seen, therefore revealing other nodes present in the network that would not have been viewed previously. This can be seen in Figure 4. What cannot be readily seen in the figure is that there are numerous singleton nodes due to removing the #metoo node. In this pathway, since the focus is on the nodes located in the margins, neither edges nor nodes are removed from the network.

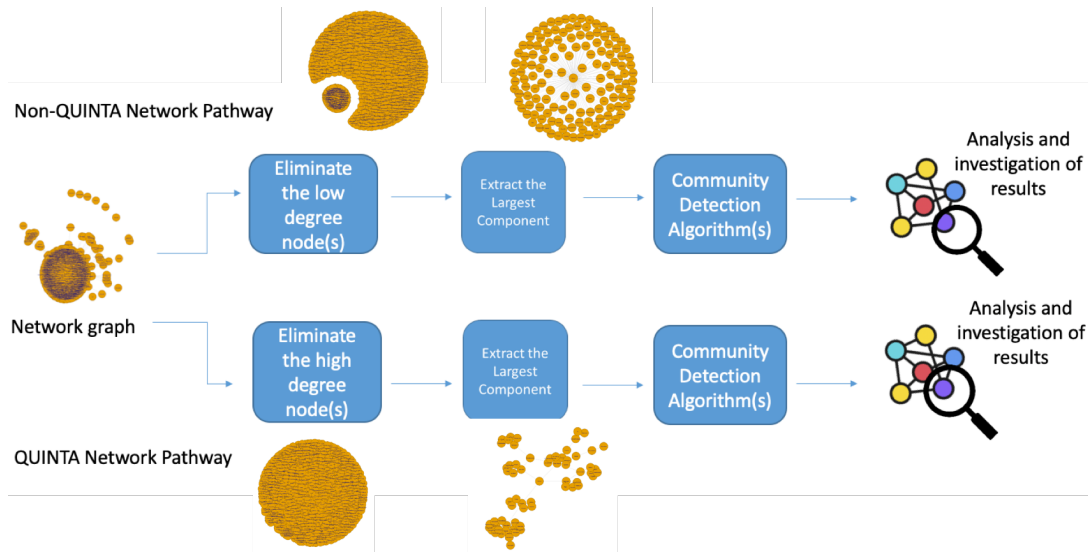

Figure 4: QUINTA Explore in #MeToo Case Study– [Upper] Non-QUINTA network path shows the method for centering on the popular discourse related to the #metoo hashtag. [Lower] QUINTA pathway illustrates the technique for identifying smaller communities beyond the mainstream discourse.

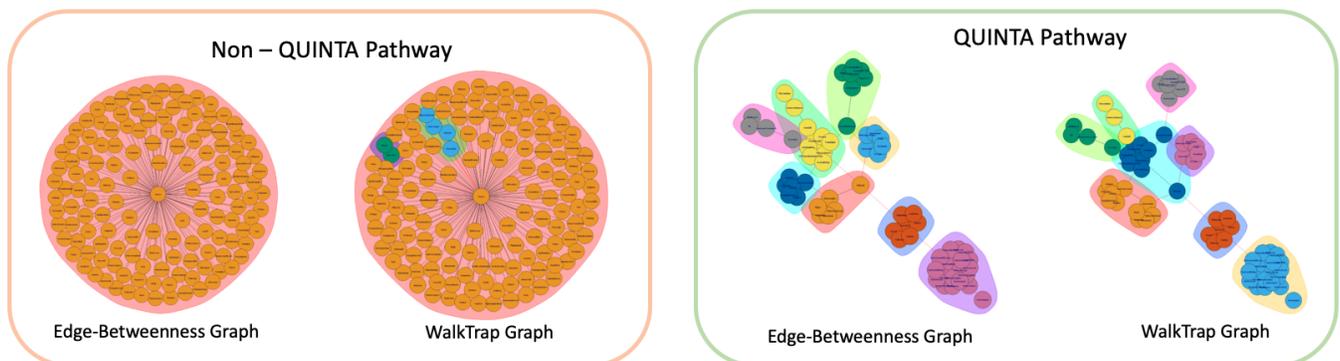

Figure 5: QUINTA Model in #MeToo Case Study– [Left] The Non-QUINTA Pathway shows the broader experiences in mass media. [Right] QUINTA Pathway shows hashtag derivatives of smaller community networks.

### 5.6 Interpret

**Are findings being interpreted to represent a universal or normative experience?** The Non-QUINTA pathway (Figure 5 [Left]) was extremely reflective of broader experiences in mass media and other public spaces. Also, we saw communities with general #metoo support especially in the Edge-Betweenness community detection algorithm. This included #timesup and a specialized community advocating for discussions about sexual assault and violence for k-12. Thus, it is easy to say that these communities are amply represented in our findings. However, in the QUINTA pathway (Figure 5 [Right]), the hashtag derivatives[6] of #metoo were more prevalent, but these communities were not as prominently discussed in the media. Examples include #metooblackchurch, #metooqueer, #metoomuslim,

[6]"Hashtag derivatives are defined as a hashtag whose composition varies on, but strongly reflects, the original hashtag." [16][p. 60]

#metoodisabled, and #ustoo, to name a few. The universal aspect of these experiences is straightforward: the ubiquity of sexual harassment and violence. By intentionally looking for these voices, these impacted communities and their experiences are valid and differ from majority voice experiences.

**Are you considering how similarities and differences in outcomes are interpreted and how the structural inequalities are enhanced by the algorithms and statistics?** Looking more granularity at clustered hashtag communities between the QUINTA and Non-QUINTA pathway, it is observed slight differences in the number of hashtags yielded, along with which hashtag that most stood out as important in the community. Notably, both community detection algorithms produced the same number of communities and nearly the same importance, even though the Non-QUINTA pathway reflected more network degree and density. This affirms

the importance of grounding a task in structural differences, reflected in a social context. Within the context of the well-known existing inequalities between these two groups of communities - roughly speaking, the dominant and the marginalized - I believe that this work helps to reduce these inequalities by centering each group of communities respectively.

**Who are you silencing/amplifying in the interpretation of your model outcomes (*e.g.,* sensitivity to nuanced variations in the data)?** It is understood through previous work [16] that the majority voice in the network analyses predominantly reflected only the #metoo hashtag. This was observed in the analysis: the Non-QUINTA pathway was highly reflective of what was happening in mass media and the conversations that were going on in public spaces, yet it was only able to reflect more homogeneous tags of #metoo. The impact of this is a silencing of non-dominant voices. In contrast, when approaching the analysis via a QUINTA Pathway, the centered tag heterogeneity is observed, resulting in a larger representation of diverse communities; these are the ones we amplify. Nonetheless, the U.S. indigenous communities are entirely missing from this conversation. Despite our best efforts at methodological inclusion, a significant gap remains that needs to be addressed in future work.

## 6 DISCUSSION

QUINTA grapples and employs reflexivity and intersectionality paradigms to question and illuminate where the power lies in the research process. This paradigm encourages researchers to eschew the prescriptive, rote habits of traditional workflows, and take steps to be equitable and inclusive. The goal of QUINTA is to begin making better decisions about the implementation of our methodologies. This means we do not engage in witch hunts that assume researchers were not or cannot be reflexive; nevertheless, it leaves us the space to critique our work and explore opportunities to be more inclusive. Furthermore, QUINTA presents a more specialized, targeted way to design equitable data pathways that more thoroughly represent inequalities. This framework also opens the door for other quantitative domains to think more critically about diversity and differences across the tech universe.

### 6.1 Limitations and Future Work

The purpose of reflexivity is to be engaged throughout the entirety of the research process. Reflexivity is not a passive concept; therefore, neither is QUINTA a passive methodology. With reflexivity, QUINTA requires conscientious engagement and intentional inclusivity. The flip side of this is that it manifests as a limitation: when not engaged with conscientiously or comprehensively, QUINTA risks becoming a fig leaf, the latest in a long line of buzzwords. An additional limitation is immediately thinking this work is only for the data science community. Even though QUINTA was created around the data science process, this process is a proxy for any other data centric processes. The example illustrated in this paper focused on network analysis. Further, Howison and colleagues [58] cautioned researchers in network analysis to be mindful of the method designs they choose for these networks impact the people they include or exclude, particularly marginalized communities.

For future research, it is encouraged that researchers (and practitioners) put this work into praxis, again echoing the intersectionality scholars [12, 28, 29]. The reflexive questions will be reframed to fit a different context or even a domain. Nevertheless, we can then build a body of knowledge, a scholarship, and fruitful conversations to begin to understand the implications and biases incorporated into the research process from the researcher's standpoint. As researchers and practitioners, we must be willing to be vulnerable in exploring the potential complexity of research topics and communities we are not a part of and may not intuitively understand. Instead of being timid, perfunctory spectators to inequality, we can get involved and use the questions posed in Table 1 to engage with the work.

## 7 CONCLUSION

Data is not neutral. The profound social, technological, and financial investments in building data infrastructures speak to its power and the need for new kinds of critiques and alternatives that center multiply-marginalized perspectives, rather than once again quixotically attempting to reform dominant defaults. This paper introduced a framework for centering AI researcher reflexivity in order to drive responsible ML and AI fairness research via the grappling of power. The framework, QUINTA, shows how reflexivity can help conceptualize and call attention to the role and decisions researchers make in this process. These decisions and model harm outcomes to historically marginalized communities, are not independent of one another. Within the QUINTA framework, reflexivity and intersectionality provides the foundation and work together. The QUINTA framework is a strong vehicle for integrating intersectional perspectives into data centric processes. There is justifiable theoretical, methodological, and epistemological tension between these perspectives, making it especially ripe for exploration and synthesis. Use reflexivity (and intersectionality) in QUINTA to critically reflect on your role (as researchers and scientists) in how your decisions (whether intentional or unintentional) harm vulnerable communities, reinforce bias, and perpetuate inequities. Implementing QUINTA comes with a disclaimer: it will take some perspective-taking, exposing one's blind spots throughout the process, and questioning the practitioner's modes of operating. QUINTA is not a feel-good approach; it is a transformative approach meant to unveil uncomfortable truths and perspectives, empowering researchers to be agents of change in their own research – starting with their-self.

## ACKNOWLEDGMENTS

This paper has been labor of love. With that, a very special thank you to Anaelia Ovalle for seeing the importance and for their valuable comments and helpful suggestions for this paper.

## REFERENCES
[1] Maria Adams. 2021. Intersectionality and reflexivity: Narratives from a BME female researcher inside the hidden social world of prison visits. *International Journal of Qualitative Methods* 20 (2021), 1609406920982141.
[2] Mats Alvesson and K Sköldberg. 2009. Reflexive Methodology. (2009).
[3] Aileen Barrett, Anu Kajamaa, and Jenny Johnston. 2020. How to… be reflexive when conducting qualitative research. *The clinical teacher* 17, 1 (2020), 9–12.
[4] Judy E Battaglia, Paige P Edley, and Victoria Ann Newsom. 2019. Intersectional feminisms and sexual violence in the era of Me Too, Trump, and Kavanaugh. *Women & Language* 42, 1 (2019), 133–143.

[5] Greta R Bauer and Daniel J Lizotte. 2021. Artificial intelligence, intersectionality, and the future of public health. , 98–100 pages.
[6] Sebastian Benthall and Bruce D. Haynes. 2019. Racial Categories in Machine Learning. In *Proceedings of the Conference on Fairness, Accountability, and Transparency* (Atlanta, GA, USA) *(FAT\* '19)*. Association for Computing Machinery, New York, NY, USA, 289–298. https://doi.org/10.1145/3287560.3287575
[7] Patrick Biernacki and Dan Waldorf. 1981. Snowball sampling: Problems and techniques of chain referral sampling. *Sociological methods & research* 10, 2 (1981), 141–163.
[8] Abeba Birhane and Olivia Guest. 2020. Towards decolonising computational sciences. *arXiv preprint arXiv:2009.14258* (2020).
[9] Abeba Birhane, William Isaac, Vinodkumar Prabhakaran, Mark Díaz, Madeleine Clare Elish, Iason Gabriel, and Shakir Mohamed. 2022. Power to the People? Opportunities and Challenges for Participatory AI. *Equity and Access in Algorithms, Mechanisms, and Optimization* (2022), 1–8.
[10] Lisa Bowleg. 2008. When Black+ lesbian+ woman≠ Black lesbian woman: The methodological challenges of qualitative and quantitative intersectionality research. *Sex roles* 59, 5-6 (2008), 312–325.
[11] Lisa Bowleg. 2012. The problem with the phrase women and minorities: intersectionality—an important theoretical framework for public health. *American journal of public health* 102, 7 (2012), 1267–1273.
[12] Lisa Bowleg. 2021. "The master's tools will never dismantle the master's house": Ten critical lessons for Black and other health equity researchers of color. *Health Education & Behavior* 48, 3 (2021), 237–249.
[13] Lisa Bowleg and Greta Bauer. 2016. Quantifying intersectionality. *Psychology of Women Quarterly* 40, 3 (2016), 337–341.
[14] AE Boyd. 2021. Intersectionality and reflexivity—decolonizing methodologies for the data science process. *Patterns* 2, 12 (2021), 100386.
[15] Alicia Boyd and Bree McEwan. 2022. Viral paradox: The intersection of "me too" and# MeToo. *New Media & Society* (2022), 14614448221099187.
[16] Alicia E. Boyd. 2021. *Quantitative intersectional data (QUINTA): a #metoo case study*. Ph. D. Dissertation. DePaul University, College of Computing and Digital Media Dissertations.
[17] Kath Browne. 2005. Snowball sampling: using social networks to research non-heterosexual women. *International journal of social research methodology* 8, 1 (2005), 47–60.
[18] Tarana Burke. 2013. The inception. http://justbeinc.wixsite.com/justbeinc/the-me-too-movement-cmml.
[19] Tarana Burke. 2021. Tarana Burke on watching MeToo go viral in 2017. https://time.com/6097392/tarana-burke-me-too-unbound-excerpt/
[20] Scott Allen Cambo and Darren Gergle. 2022. Model Positionality and Computational Reflexivity: Promoting Reflexivity in Data Science. In *CHI Conference on Human Factors in Computing Systems*. 1–19.
[21] Chris Chapman. 2011. Resonance, intersectionality, and reflexivity in critical pedagogy (and research methodology). *Social Work Education* 30, 7 (2011), 723–744.
[22] Devika Chawla. 2019. My #MeToos before the #MeToo. *Women & Language* 42 (2019), 165–168.
[23] Andrew JG Cockburn and Harold Thimbleby. 1991. A reflexive perspective of CSCW. *ACM SIGCHI Bulletin* 23, 3 (1991), 63–68.
[24] Elizabeth R Cole. 2009. Intersectionality and research in psychology. *American psychologist* 64, 3 (2009), 170.
[25] Combahee River Collective. 1983. The Combahee river collective statement. *Home girls: A Black feminist anthology* (1983), 264–74.
[26] Patricia Hill Collins. 1991. *Black feminist thought: Knowledge, consciousness, and the politics of empowerment*. routledge.
[27] Patricia Hill Collins. 2015. Intersectionality's Definitional Dilemmas. *Annual Review of Sociology* 41, 1 (2015), 1–20. https://doi.org/10.1146/annurev-soc-073014-112142
[28] Patricia Hill Collins. 2019. *Intersectionality as critical social theory*. Duke University Press.
[29] Patricia Hill Collins and Sirma Bilge. 2016. *Intersectionality*. John Wiley & Sons.
[30] Kimberle Crenshaw. 1989. Demarginalizing the intersection of race and sex: A black feminist critique of antidiscrimination doctrine, feminist theory and anti-racist politics. *u. Chi. Legal f.* (1989), 139.
[31] Kimberle Crenshaw. 1990. Mapping the margins: Intersectionality, identity politics, and violence against women of color. *Stan. L. Rev.* 43 (1990), 1241.
[32] Wesam Darawsheh. 2014. Reflexivity in research: Promoting rigour, reliability and validity in qualitative research. *International journal of therapy and rehabilitation* 21, 12 (2014), 560–568.
[33] Cécile Deer. 2014. Reflexivity. In *Pierre Bourdieu*. Routledge, 207–220.
[34] Catherine D'ignazio and Lauren F Klein. 2020. *Data feminism*. MIT press.
[35] Cristy Dougherty and Bernadette Marie Calafell. 2019. Before and Beyond #MeToo and #TimesUp: Rape as a Colonial and Racist Project. *Women & Language* 42 (2019), 181–185.
[36] Maura Dowling. 2006. Approaches to reflexivity in qualitative research. *Nurse researcher* 13, 3 (2006).
[37] Margaret U D'silva, Siobhan E Smith, Lindsay J Della, Deborah A Potter, Theresa A Rajack-Talley, and Latrica Best. 2016. Reflexivity and Positionality in Researching African-American Communities: Lessons from the Field. *Intercultural Communication Studies* 25, 1 (2016).
[38] Jane Elliott. 2005. The researcher as narrator: reflexivity in qualitative and quantitative research. *Using narrative in social research: qualitative and quantitative approaches* (2005), 152–159.
[39] Nicole M. Else-Quest and Janet Shibley Hyde. 2016. Intersectionality in Quantitative Psychological Research: I. Theoretical and Epistemological Issues. *Psychology of Women Quarterly* 40, 2 (2016), 155–170. https://doi.org/10.1177/0361684316629797
[40] Nicole M. Else-Quest and Janet Shibley Hyde. 2016. Intersectionality in Quantitative Psychological Research: II. Methods and Techniques. *Psychology of Women Quarterly* 40, 3 (2016), 319–336. https://doi.org/10.1177/0361684316647953
[41] Linda Finlay. 2002. Negotiating the swamp: the opportunity and challenge of reflexivity in research practice. *Qualitative research* 2, 2 (2002), 209–230.
[42] Benjamin Fish and Luke Stark. 2021. Reflexive design for fairness and other human values in formal models. In *Proceedings of the 2021 AAAI/ACM Conference on AI, Ethics, and Society*. 89–99.
[43] Barton Friedland and Yutaka Yamauchi. 2011. Reflexive design thinking: putting more human in human-centered practices. *interactions* 18, 2 (2011), 66–71.
[44] Kay Fuller. 2020. The "7 up" intersectionality life grid: A tool for reflexive practice. In *Frontiers in Education*, Vol. 5. Frontiers Media SA, 77.
[45] Simson Garfinkel, Jeanna Matthews, Stuart S Shapiro, and Jonathan M Smith. 2017. Toward algorithmic transparency and accountability. , 5–5 pages.
[46] Timnit Gebru, Jamie Morgenstern, Briana Vecchione, Jennifer Wortman Vaughan, Hanna Wallach, Hal Daumé Iii, and Kate Crawford. 2021. Datasheets for datasets. *Commun. ACM* 64, 12 (2021), 86–92.
[47] Marco Gemignani and Yolanda Hernández-Albújar. 2019. Critical reflexivity and intersectionality in human rights. *European Psychologist* (2019).
[48] Meghan Gilbert-Hickey. 2019. #MeToo, Moving Forward: How Reckoning with an Imperfect Movement Can Help Us Examine Violent Inequality, Past and Present, In Order to Dismantle it in the Future. *South Central Review* 36, 2 (2019), 1–16.
[49] Rosalind Gill and Shani Orgad. 2018. The shifting terrain of sex and power: From the 'sexualization of culture' to# MeToo. *Sexualities* 21, 8 (2018), 1313–1324.
[50] Jennifer M Gómez and Robyn L Gobin. 2020. Black women and girls & #MeToo: Rape, cultural betrayal, & healing. *Sex Roles* 82, 1-2 (2020), 1–12.
[51] Breda Gray. 2008. Putting emotion and reflexivity to work in researching migration. *Sociology* 42, 5 (2008), 935–952.
[52] Paul Greenbank. 2003. The role of values in educational research: The case for reflexivity. *British educational research journal* 29, 6 (2003), 791–801.
[53] Marilys Guillemin and Lynn Gillam. 2004. Ethics, reflexivity, and "ethically important moments" in research. *Qualitative inquiry* 10, 2 (2004), 261–280.
[54] Lelia Marie Hampton. 2021. Black Feminist Musings on Algorithmic Oppression. *arXiv preprint arXiv:2101.09869* (2021).
[55] Helen Hand. 2003. The mentor's tale: a reflexive account of semi-structured interviews. *Nurse Researcher (through 2013)* 10, 3 (2003), 15.
[56] Karen Henwood. 2008. Qualitative research, reflexivity and living with risk: Valuing and practicing epistemic reflexivity and centering marginality. *Qualitative research in psychology* 5, 1 (2008), 45–55.
[57] Zahara Hill. 2017. A Black Woman Created the "Me Too" Campaign Against Sexual Assault 10 Years Ago. (2017). https://www.ebony.com/news/black-woman-me-too-movement-tarana-burke-alyssa-milano/
[58] James Howison, Andrea Wiggins, and Kevin Crowston. 2011. Validity issues in the use of social network analysis with digital trace data. *Journal of the Association for Information Systems* 12, 12 (2011), 2.
[59] V Jo Hsu. 2019. (Trans) forming #MeToo: Toward a Networked Response to Gender Violence. *Women's Studies in Communication* 42, 3 (2019), 269–286.
[60] Alina Huldtgren and Cordula Endter. 2014. Reflexive practice in interdisciplinary design of pervasive health applications in dementia care. In *Proceedings of the 8th International Conference on Pervasive Computing Technologies for Healthcare*. 244–247.
[61] Tracey L Hurd. 1998. Process, content, and feminist reflexivity: One researcher's exploration. *Journal of Adult Development* 5, 3 (1998), 195–203.
[62] Ben Hutchinson and Margaret Mitchell. 2019. 50 Years of Test (Un)Fairness: Lessons for Machine Learning. In *Proceedings of the Conference on Fairness, Accountability, and Transparency* (Atlanta, GA, USA) *(FAT\* '19)*. Association for Computing Machinery, New York, NY, USA, 49–58. https://doi.org/10.1145/3287560.3287600
[63] Richard G Jones. 2010. Putting privilege into practice through" intersectional reflexivity:" Ruminations, interventions, and possibilities. *Reflections: Narratives of Professional Helping* (2010), 122–125.
[64] Jane Jorgenson. 2011. Reflexivity in feminist research practice: Hearing the unsaid. *Women & Language* 34, 2 (2011), 115.
[65] Os Keyes. 2019. Counting the Countless: Why data science is a profound threat for queer people. *Real Life* 2 (2019).

[66] Cat LaFuente. 2017. Who is the woman behind the #MeToo Movement? https://www.thelist.com/110186/woman-behind-metoo-movement/. (2017).
[67] Heather Lang. 2019. #MeToo: A Case Study in Re-Embodying Information. *Computers and Composition* 53 (2019), 9–20.
[68] Michael Lynch. 2000. Against reflexivity as an academic virtue and source of privileged knowledge. *Theory, Culture & Society* 17, 3 (2000), 26–54.
[69] Douglas Macbeth. 2001. On "reflexivity" in qualitative research: Two readings, and a third. *Qualitative inquiry* 7, 1 (2001), 35–68.
[70] Ashley Noel Mack and Bryan J McCann. 2018. Critiquing state and gendered violence in the age of #MeToo. *Quarterly Journal of Speech* 104, 3 (2018), 329–344.
[71] Samantha Majic and Carisa R Showden. 2018. Redesigning the study of sex work: A case for intersectionality and reflexivity. In *Routledge International Handbook of Sex Industry Research*. Routledge, 42–54.
[72] Lydia Manikonda, Ghazaleh Beigi, Subbarao Kambhampati, and Huan Liu. 2018. #metoo Through the Lens of Social Media. In *International Conference on Social Computing, Behavioral-Cultural Modeling and Prediction and Behavior Representation in Modeling and Simulation*. Springer, 104–110.
[73] Lydia Manikonda, Ghazaleh Beigi, Huan Liu, and Subbarao Kambhampati. 2018. Twitter for Sparking a Movement, Reddit for Sharing the Moment:# metoo through the Lens of Social Media. *arXiv preprint arXiv:1803.08022* (2018).
[74] Jennifer Mason. 1996. Qualitative researching. (1996).
[75] Denys JC Matthies, Bodo Urban, Katrin Wolf, and Albrecht Schmidt. 2019. Reflexive Interaction: Extending the concept of Peripheral Interaction. In *Proceedings of the 31st Australian Conference on Human-Computer-Interaction*. 266–278.
[76] R Maule. 2020. Not just a movement for famous white cisgendered women. *Me Too and intersectionality. Gender and Women's Studies* 2, 3 (2020), 1–13.
[77] Natasha S Mauthner and Andrea Doucet. 2003. Reflexive accounts and accounts of reflexivity in qualitative data analysis. *Sociology* 37, 3 (2003), 413–431.
[78] Mary Maynard. 2013. Methods, practice and epistemology: The debate about feminism and research. In *Researching women's lives from a feminist perspective*. Routledge, 10–26.
[79] Milagros Miceli, Tianling Yang, Laurens Naudts, Martin Schuessler, Diana Serbanescu, and Alex Hanna. 2021. Documenting computer vision datasets: an invitation to reflexive data practices. In *Proceedings of the 2021 ACM Conference on Fairness, Accountability, and Transparency*. 161–172.
[80] Margaret Mitchell, Simone Wu, Andrew Zaldivar, Parker Barnes, Lucy Vasserman, Ben Hutchinson, Elena Spitzer, Inioluwa Deborah Raji, and Timnit Gebru. 2019. Model cards for model reporting. In *Proceedings of the conference on fairness, accountability, and transparency*. 220–229.
[81] Saki Mizoroki, Limor Shifman, and Kaori Hayashi. 2023. Hashtag activism found in translation: Unpacking the reformulation of# MeToo in Japan. *new media & society* (2023), 14614448231153571.
[82] Deirdre K. Mulligan, Joshua A. Kroll, Nitin Kohli, and Richmond Y. Wong. 2019. This Thing Called Fairness: Disciplinary Confusion Realizing a Value in Technology. *Proc. ACM Human Computer Interact.* 3, CSCW, Article 119 (Nov. 2019), 36 pages. https://doi.org/10.1145/3359221
[83] Rory V O'Connor and Paul Clarke. 2015. Software process reflexivity and business performance: initial results from an empirical study. In *Proceedings of the 2015 International Conference on Software and System Process*. 142–146.
[84] Anna Oleszczuk et al. 2020. #Hashtag: How Selected Texts of Popular Culture Engaged With Sexual Assault In the Context of the Me Too Movement in 2019. *New Horizons in English Studies* 5, 1 (2020), 208–217.
[85] Cathy O'Neil. 2016. *Weapons of math destruction: How big data increases inequality and threatens democracy*. Crown.
[86] Angela Onwuachi-Willig. 2018. What about#UsToo: the invisibility of race in the#MeToo movement. *Yale LJF* 128 (2018), 105.
[87] Anaelia Ovalle, Arjun Subramonian, Vagrant Gautam, Gilbert Gee, and Kai-Wei Chang. 2023. Factoring the Matrix of Domination: A Critical Review and Reimagination of Intersectionality in AI Fairness. *arXiv preprint arXiv:2303.17555* (2023).
[88] Yushan Pan. 2021. Reflexivity of Account, Professional Vision, and Computer-Supported Cooperative Work: Working in the Maritime Domain. *Proceedings of the ACM on Human-Computer Interaction* 5, CSCW2 (2021), 1–32.
[89] Louise Phillips, Marianne Kristiansen, Marja Vehviläinen, and Ewa Gunnarson. 2013. Tackling the tensions of dialogue and participation. *Knowledge and power in collaborative research: A reflexive approach* 6 (2013), 1.
[90] Alison Phipps. 2019. " Every Woman Knows a Weinstein": Political Whiteness and White Woundedness in# MeToo and Public Feminisms around Sexual Violence. *Feminist Formations* 31, 2 (2019), 1–25.
[91] Suvi Pihkala and Helena Karasti. 2016. Reflexive engagement: enacting reflexivity in design and for'participation in plural'. In *Proceedings of the 14th Participatory Design Conference: Full Papers-Volume 1*. 21–30.
[92] Wanda Pillow. 2003. Confession, catharsis, or cure? Rethinking the uses of reflexivity as methodological power in qualitative research. *International journal of qualitative studies in education* 16, 2 (2003), 175–196.
[93] Amon Rapp. 2018. Reflexive ethnographies in human-computer interaction: Theory and practice. In *Extended Abstracts of the 2018 CHI Conference on Human Factors in Computing Systems*. 1–4.
[94] Carla Rice, Elisabeth Harrison, and May Friedman. 2019. Doing justice to intersectionality in research. *Cultural Studies Critical Methodologies* 19, 6 (2019), 409–420.
[95] P Roche, C Shimmin, S Hickes, M Khan, O Sherzoi, E Wicklund, JG Lavoie, S Hardie, KDM Wittmeier, and KM Sibley. 2020. Valuing All Voices: refining a trauma-informed, intersectional and critical reflexive framework for patient engagement in health research using a qualitative descriptive approach. *Research involvement and engagement* 6, 1 (2020), 1–13.
[96] Jennifer A Rode. 2011. Reflexivity in digital anthropology. In *Proceedings of the SIGCHI conference on human factors in computing systems*. 123–132.
[97] Bonnie Ruberg and Spencer Ruelos. 2020. Data for queer lives: How LGBTQ gender and sexuality identities challenge norms of demographics. *Big Data & Society* 7, 1 (2020), 2053951720933286.
[98] Louise Ryan and Anne Golden. 2006. 'Tick the box please': A reflexive approach to doing quantitative social research. *Sociology* 40, 6 (2006), 1191–1200.
[99] Tess Ryan. 2019. For Indigenous women, #MeToo is a fight against racism and oppression. https://www.sbs.com.au/nitv/nitv-news/article/2019/10/28/indigenous-women-metoo-fight-against-racism-and-oppression
[100] Sara Salem. 2018. Intersectionality and its discontents: Intersectionality as traveling theory. *European Journal of Women's Studies* 25, 4 (2018), 403–418.
[101] Philip Carl Salzman. 2002. On reflexivity. *American Anthropologist* 104, 3 (2002), 805–811.
[102] Andrew D. Selbst, Danah Boyd, Sorelle A. Friedler, Suresh Venkatasubramanian, and Janet Vertesi. 2019. Fairness and Abstraction in Sociotechnical Systems. In *Proceedings of the Conference on Fairness, Accountability, and Transparency* (Atlanta, GA, USA) *(FAT\* '19)*. Association for Computing Machinery, New York, NY, USA, 59–68. https://doi.org/10.1145/3287560.3287598
[103] Phoebe Sengers, Kirsten Boehner, Shay David, and Joseph'Jofish' Kaye. 2005. Reflective design. In *Proceedings of the 4th decennial conference on Critical computing: between sense and sensibility*. 49–58.
[104] Rozina Sini. 2017. How 'MeToo' is exposing the scale of sexual abuse. https://www.bbc.com/news/blogs-trending-41633857
[105] Susan Smith. 2006. Encouraging the use of reflexivity in the writing up of qualitative research. *International Journal of Therapy and Rehabilitation* 13, 5 (2006), 209–215.
[106] Jiyoun Suk, Aman Abhishek, Yini Zhang, So Yun Ahn, Teresa Correa, Christine Garlough, and Dhavan V Shah. 2019. #MeToo, Networked Acknowledgment, and Connective Action: How "Empowerment Through Empathy" Launched a Social Movement. *Social Science Computer Review* (2019), 0894439319864882.
[107] Farhana Sultana. 2007. Reflexivity, positionality and participatory ethics: Negotiating fieldwork dilemmas in international research. *ACME: An international journal for critical geographies* 6, 3 (2007), 374–385.
[108] Harini Suresh, Rajiv Movva, Amelia Lee Dogan, Rahul Bhargava, Isadora Cruxen, Ángeles Martinez Cuba, Guilia Taurino, Wonyoung So, and Catherine D'Ignazio. 2022. Towards Intersectional Feminist and Participatory ML: A Case Study in Supporting Feminicide Counterdata Collection. In *2022 ACM Conference on Fairness, Accountability, and Transparency*. 667–678.
[109] Ashwini Tambe. 2018. Reckoning with the Silences of #MeToo. *Feminist Studies* 44, 1 (2018), 197–203.
[110] Toniesha L. Taylor. 2019. Dear Nice White Ladies: A Womanist Response to Intersectional Feminism and Sexual Violence. *Women & Language* 42 (2019), 187–190.
[111] Verity Trott. 2020. Networked feminism: counterpublics and the intersectional issues of# MeToo. *Feminist Media Studies* (2020), 1–18.
[112] Susan Walker, Susan Read, and Helena Priest. 2013. Use of reflexivity in a mixed-methods study. *Nurse researcher* 20, 3 (2013).
[113] Ron Weber. 2003. Editor's comments: The reflexive researcher. *MIS Quarterly* (2003), v–xiv.
[114] Joseph Webster. 2008. Establishing the 'truth'of the matter: Confessional reflexivity as introspection and avowal. *Psychology and Society* 1, 1 (2008), 65–76.
[115] Ying Xiong, Moonhee Cho, and Brandon Boatwright. 2019. Hashtag activism and message frames among social movement organizations: Semantic network analysis and thematic analysis of Twitter during the #MeToo movement. *Public Relations Review* 45, 1 (2019), 10–23.
[116] Zhao Yang, René Algesheimer, and Claudio J Tessone. 2016. A comparative analysis of community detection algorithms on artificial networks. *Scientific reports* 6 (2016), 30750.
[117] Dubravka Zarkov and Kathy Davis. 2018. Ambiguities and dilemmas around #Metoo: #forhow long and #whereto?